\documentclass[twocolumn]{aastex63}
\usepackage{graphicx}
\usepackage{ulem}
\usepackage{cellspace}
\usepackage{amssymb}
\usepackage{amsmath}
\usepackage{longtable}
\usepackage{tabu}
\usepackage{color}
\usepackage{etoolbox}
\usepackage{supertabular}
\usepackage{savesym}
\savesymbol{tablenum}
\usepackage{siunitx}
\restoresymbol{SIX}{tablenum}
\usepackage{dcolumn}
\usepackage{ wasysym }

\newcommand\Tstrut{\rule{0pt}{2.9ex}}       
\newcommand\Bstrut{\rule[-1.3ex]{0pt}{0pt}} 
\newcommand\TBstrut{\Tstrut\Bstrut}  

\newcommand{\dataStartGPS}{\num{1238166018}}
\newcommand{\dataStartDate}{2019 April 1 15:00:00 GMT}
\newcommand{\dataEndGPS}{\num{1254150018}}
\newcommand{\dataEndDate}{2019 October 3 15:00:00 GMT}
\newcommand{\paramtotaltemplates}{\num{6.7e+18}}
\newcommand{\paramtotalWUsmillions}{\num{55.7}}
\newcommand{\nCandPerWUHalfHz}{\num{30000}}
\newcommand{\nCandPerWUfiftymHz}{\num{3000}}
\newcommand{\paramNCandTotal}{\num{1.7e+12}}
\newcommand{\paramWUtotaltemplates}{\num{1.2e+11}}
\newcommand{\paramfmax}{800.0}
\newcommand{\paramfmin}{20.0}
\newcommand{\paramfdotmin}{-2.6e-9}
\newcommand{\paramfdotmax}{2.6e-10}
\newcommand{\paramTref}{1246070525.0}
\newcommand{\TcohFUZero}{\num{120}}
\newcommand{\dfreqmuHzFUZero}{\num{2}}
\newcommand{\dfdotfgTenfHzFUZero}{\num{60}}
\newcommand{\mskyFUZero}{\num{0.002}}

\newcommand{\dfreqmuHzFUOne}{\num{1}}
\newcommand{\dfdotfgTenfHzFUOne}{\num{15}}
\newcommand{\mskyFUOne}{\num{0.0002}}
\newcommand{\TcohFUOne}{\num{120}}
\newcommand{\SRfreqmuHzFUOne}{\num{1000}}
\newcommand{\SRfdotFUOne}{\num{11250}}
\newcommand{\SRskyFUOne}{\num{10.0}}
\newcommand{\rVetoFUOne}{-}
\newcommand{\aTwoFrFUOne}{8.5}
\newcommand{\aBSGLtLrFUOne}{-1}
\newcommand{\dfreqmuHzFUTwo}{\num{1}}
\newcommand{\dfdotfgTenfHzFUTwo}{\num{2}}
\newcommand{\mskyFUTwo}{\num{2e-6}}
\newcommand{\TcohFUTwo}{\num{120}}
\newcommand{\SRfreqmuHzFUTwo}{\num{50}}
\newcommand{\SRfdotFUTwo}{\num{1200}}
\newcommand{\SRskyFUTwo}{\num{1.0}}
\newcommand{\rVetoFUTwo}{-}
\newcommand{\aTwoFrFUTwo}{10}
\newcommand{\aBSGLtLrFUTwo}{-1}
\newcommand{\dfreqmuHzFUThree}{\num{0.1}}
\newcommand{\dfdotfgTenfHzFUThree}{\num{0.2}}
\newcommand{\mskyFUThree}{\num{1.0e-8}}
\newcommand{\TcohFUThree}{\num{240}}
\newcommand{\SRfreqmuHzFUThree}{\num{5}}
\newcommand{\SRfdotFUThree}{\num{200}}
\newcommand{\SRskyFUThree}{\num{0.2}}
\newcommand{\rVetoFUThree}{-}
\newcommand{\aTwoFrFUThree}{14}
\newcommand{\aBSGLtLrFUThree}{-5}
\newcommand{\dfreqmuHzFUFour}{\num{0.013}}
\newcommand{\dfdotfgTenfHzFUFour}{\num{0.064}}
\newcommand{\mskyFUFour}{\num{3.2e-10}}
\newcommand{\TcohFUFour}{\num{490}}
\newcommand{\SRfreqmuHzFUFour}{\num{0.5}}
\newcommand{\SRfdotFUFour}{\num{35}}
\newcommand{\SRskyFUFour}{\num{0.04}}
\newcommand{\rVetoFUFour}{3}
\newcommand{\aTwoFrFUFour}{-}
\newcommand{\aBSGLtLrFUFour}{-}
\newcommand{\dfreqmuHzFUFive}{\num{0.001}}
\newcommand{\dfdotfgTenfHzFUFive}{\num{0.032}}
\newcommand{\mskyFUFive}{\num{1.4e-10}}
\newcommand{\TcohFUFive}{\num{1100}}
\newcommand{\SRfreqmuHzFUFive}{\num{0.2}}
\newcommand{\SRfdotFUFive}{\num{20}}
\newcommand{\SRskyFUFive}{\num{0.017}}
\newcommand{\rVetoFUFive}{7}
\newcommand{\aTwoFrFUFive}{-}
\newcommand{\aBSGLtLrFUFive}{-}
\newcommand{\dfreqmuHzFUSix}{\num{0.001}}
\newcommand{\dfdotfgTenfHzFUSix}{\num{0.009}}
\newcommand{\mskyFUSix}{\num{2.8e-11}}
\newcommand{\TcohFUSix}{\num{2200}}
\newcommand{\SRfreqmuHzFUSix}{\num{0.1}}
\newcommand{\SRfdotFUSix}{\num{8.5}}
\newcommand{\SRskyFUSix}{\num{0.008}}
\newcommand{\rVetoFUSix}{15}
\newcommand{\aTwoFrFUSix}{-}
\newcommand{\aBSGLtLrFUSix}{-}
\newcommand{\dfreqmuHzFUSeven}{\num{0.001}}
\newcommand{\dfdotfgTenfHzFUSeven}{\num{0.0063}}
\newcommand{\mskyFUSeven}{\num{1e-11}}
\newcommand{\TcohFUSeven}{\mathrm{coherent}}
\newcommand{\SRfreqmuHzFUSeven}{\num{0.07}}
\newcommand{\SRfdotFUSeven}{\num{6}}
\newcommand{\SRskyFUSeven}{\num{0.0057}}
\newcommand{\rVetoFUSeven}{35}
\newcommand{\aTwoFrFUSeven}{-}
\newcommand{\aBSGLtLrFUSeven}{-}
\newcommand{\dfreqmuHzFUEight}{\num{0.001}}
\newcommand{\dfdotfgTenfHzFUEight}{\num{0.0063}}
\newcommand{\mskyFUEight}{\num{1e-11}}

\newcommand{\SRfreqmuHzFUEight}{\num{0.07}}
\newcommand{\SRfdotFUEight}{\num{6}}
\newcommand{\SRskyFUEight}{\num{0.0057}}
\newcommand{\rVetoFUEight}{15}
\newcommand{\aTwoFrFUEight}{-}
\newcommand{\aBSGLtLrFUEight}{-}
\newcommand{\dfreqmuHzFUNine}{\num{0.001}}
\newcommand{\dfdotfgTenfHzFUNine}{\num{0.0063}}
\newcommand{\mskyFUNine}{\num{1e-11}}

\newcommand{\SRfreqmuHzFUNine}{\num{0.07}}
\newcommand{\SRfdotFUNine}{\num{6}}
\newcommand{\SRskyFUNine}{\num{0.0057}}

\newcommand{\NSegFUZero}{37}
\newcommand{\NSegFUOne}{37}
\newcommand{\NSegFUTwo}{37}
\newcommand{\NSegFUThree}{19}
\newcommand{\NSegFUFour}{9}
\newcommand{\NSegFUFive}{4}
\newcommand{\NSegFUSix}{2}
\newcommand{\NSegFUSeven}{1}
\newcommand{\NSegFUEight}{1}
\newcommand{\NSegFUNine}{1}

\newcommand{\meanMismatchFUZero}{\num{56}}
\newcommand{\meanMismatchFUOne}{\num{31}}
\newcommand{\meanMismatchFUTwo}{\num{2.5}}
\newcommand{\meanMismatchFUThree}{\num{0.06}}
\newcommand{\meanMismatchFUFour}{<\num{0.01}}
\newcommand{\meanMismatchFUFive}{<\num{0.01}}
\newcommand{\meanMismatchFUSix}{<\num{0.01}}
\newcommand{\meanMismatchFUSeven}{<\num{0.01}}
\newcommand{\meanMismatchFUEight}{<\num{0.01}}
\newcommand{\meanMismatchFUNine}{<\num{0.01}}

\newcommand{\NCandFUOne}{\num{3513855}}
\newcommand{\NCandFUTwo}{\num{386429}}
\newcommand{\NCandFUThree}{\num{35635}}
\newcommand{\NCandFUFour}{\num{5116}}
\newcommand{\NCandFUFive}{\num{1387}}
\newcommand{\NCandFUSix}{\num{310}}
\newcommand{\NCandFUSeven}{\num{54}}
\newcommand{\NCandFUEight}{\num{12}}
\newcommand{\NCandFUNine}{\num{6}}
\newcommand{\noFUBands}{\num{57}}

\newcommand{\injRecoveredApprox}{\approx \num{1600}}

\newcommand{\mostStringentUL}{\num{8.1e-26}}
\newcommand{\mostStringentULFreq}{\num{203}}

\def\Tcoh{T_{\textrm{\mbox{\tiny{coh}}}}}

\def\min{\textrm{\mbox{\tiny{min}}}}
\def\max{\textrm{\mbox{\tiny{max}}}}

\def\EaH{Einstein@Home}

\renewcommand{\sec}{\mathrm{s}}

\newcommand{\avgSeg}[1]{\overline{#1}}			

\newcommand{\Freq}{f}
\newcommand{\fdot}{{\dot{\Freq}}}

\newcommand{\cosi}{\cos\iota}

\newcommand{\M}{\mathcal{M}}

\newcommand{\Gauss}{\mathrm{\MakeUppercase{G}}}
\newcommand{\Signal}{{\mathrm{\MakeUppercase{S}}}}
\newcommand{\Line}{{\mathrm{\MakeUppercase{L}}}}
\newcommand{\Transient}{{\mathrm{t\MakeUppercase{L}}}}

\newcommand{\NoisetL}{{\Gauss\Line\Transient}}

\providecommand{\sc}[1]{\widehat{#1}}
\renewcommand{\sc}[1]{\widehat{#1}}

\newcommand{\bsgltlr}{{\hat\beta}_{{\Signal/\NoisetL r}}}	
\newcommand{\BSNtsc}{{\hat\beta}_{{\Signal/\NoisetL}}}	

\newcommand{\bsgltl}{\BSNtsc}

\newcommand{\F}{\mathcal{F}}		

\newcommand{\avF}{\avgSeg{\F}}

\newcommand{\Nseg}{{N_{\mathrm{seg}}}}

\renewcommand{\csc}{$\sc{\textrm{c}}$}

\newcommand{\Depth}{{\mathcal{D}}}

\newcommand{\halfHzband}{\SI{0.5}{\hertz}}
\newcommand{\fiftyMHzband}{\SI{50}{\milli\hertz}}

\newcommand{\TwoFrThreshold}{2\avF_r^{\textrm{thr}}}
\newcommand{\BSGLtLrThreshold}{\bsgltlr^{\textrm{thr}}}

\shorttitle{Deep O3 Einstein@Home all-sky search for continuous gravitational waves}
\shortauthors{Steltner et al.}

\begin{document}

\title{Deep Einstein@Home all-sky search for continuous gravitational waves in LIGO O3 public data}

\correspondingauthor{B. Steltner}
\email{benjamin.steltner@aei.mpg.de}

\correspondingauthor{M.A. Papa}
\email{maria.alessandra.papa@aei.mpg.de}

\author[0000-0003-1833-5493]{B. Steltner}
\affiliation{Max Planck Institute for Gravitational Physics (Albert Einstein Institute), Callinstrasse 38, D-30167 Hannover, Germany}
\affiliation{Leibniz Universit\"at Hannover, D-30167 Hannover, Germany}

\author[0000-0002-1007-5298]{M. A. Papa}
\affiliation{Max Planck Institute for Gravitational Physics (Albert Einstein Institute), Callinstrasse 38, D-30167 Hannover, Germany}
\affiliation{Leibniz Universit\"at Hannover, D-30167 Hannover, Germany}

\author[0000-0001-5296-7035]{H.-B. Eggenstein}
\affiliation{Max Planck Institute for Gravitational Physics (Albert Einstein Institute), Callinstrasse 38, D-30167 Hannover, Germany}
\affiliation{Leibniz Universit\"at Hannover, D-30167 Hannover, Germany}

\author[0000-0002-3789-6424]{R. Prix}
\affiliation{Max Planck Institute for Gravitational Physics (Albert Einstein Institute), Callinstrasse 38, D-30167 Hannover, Germany}
\affiliation{Leibniz Universit\"at Hannover, D-30167 Hannover, Germany}

\author{M. Bensch}
\affiliation{Max Planck Institute for Gravitational Physics (Albert Einstein Institute), Callinstrasse 38, D-30167 Hannover, Germany}
\affiliation{Leibniz Universit\"at Hannover, D-30167 Hannover, Germany}

\author[0000-0003-4285-6256]{B. Allen}
\affiliation{Max Planck Institute for Gravitational Physics (Albert Einstein Institute), Callinstrasse 38, D-30167 Hannover, Germany}
\affiliation{Leibniz Universit\"at Hannover, D-30167 Hannover, Germany}

\author{B. Machenschalk}
\affiliation{Max Planck Institute for Gravitational Physics (Albert Einstein Institute), Callinstrasse 38, D-30167 Hannover, Germany}
\affiliation{Leibniz Universit\"at Hannover, D-30167 Hannover, Germany}

\begin{abstract}
	We present the results of an all-sky search for continuous gravitational waves in the public LIGO O3 data. The search covers signal frequencies $\SI{\paramfmin}{\hertz} \leq f \leq \SI{\paramfmax}{\hertz}$ and a spin-down range down to $-2.6\times 10^{-9}$ Hz s$^{-1}$, motivated by detectability studies on synthetic populations of Galactic neutron stars. This search is the most sensitive all-sky search to date in this frequency/spin-down region. The initial search was performed using the first half of the public LIGO O3 data (O3a), utilizing Graphical Processing Units provided in equal parts by the volunteers of the \EaH{} computing project and by the ATLAS cluster. After a hierarchical follow-up in seven stages, \NCandFUEight{} candidates remain. 
Six are discarded at the eighth stage, by using the remaining O3 LIGO data (O3b). The surviving six can be ascribed to continuous-wave fake signals present in the LIGO data for validation purposes. We recover these fake signals with very high accuracy with our last stage search, which coherently combines all O3 data. Based on our results, we set upper limits on the gravitational-wave amplitude $h_0$ and translate these in upper limits on the neutron star ellipticity and on the $r$-mode amplitude. The most stringent upper limits are at $\mostStringentULFreq$ Hz, with $h_0=\mostStringentUL$ at the 90\% confidence level. Our results exclude isolated neutron stars rotating faster than $\SI{5}{\milli\second}$ with ellipticities greater than $\num{5e-8} \left[{d\over{100~\textrm{pc}}}\right]$ within a distance $d$ from Earth and $r$-mode amplitudes $\alpha \geq 10^{-5} \left[{d\over{100~\textrm{pc}}}\right]$ for neutron stars spinning faster than $\SI{150}{\hertz}$.
\end{abstract}


\section{Introduction}
Continuous gravitational waves are nearly monochromatic, long-lasting signals. They could come from fast-rotating nonaxisymmetric neutron stars, from the excitation of unstable $r$-modes \citep{Owen:1998xg,Lasky:2015uia}, the fast inspiral of dark matter objects \citep{Horowitz:2019aim,Horowitz:2019pru} or superradiant emission of axion-like particles around back holes \citep{Arvanitaki:2014wva,Zhu:2020tht}.

The detection of a continuous gravitational wave is still elusive. Compared to the already detected gravitational waves of compact binary coalescences \citep{LIGOScientific:2018mvr,LIGOScientific:2020ibl,LIGOScientific:2021djp,Nitz:2018imz,Nitz:2020oeq,Nitz:2021uxj,Nitz:2021zwj,Venumadhav:2019tad,Venumadhav:2019lyq,Olsen:2022pin}, the continuous gravitational-wave amplitude at Earth is orders of magnitude smaller. However, since the signal is long-lasting,  one can integrate it over many months and increase the signal-to-noise ratio. 

When the waveform parameters are not known, broad parameter searches are carried out, and they are expensive because the number of waveforms that can be resolved over many months of observational data is extremely large.

In this paper, we present an all-sky search for unknown, isolated neutron stars with a gravitational-wave frequency $\SI{\paramfmin}{\hertz} \leq f \leq \SI{\paramfmax}{\hertz}$ and spin-down $\SI{\paramfdotmin}{\hertz\per\second} \leq \fdot \leq \SI{\paramfdotmax}{\hertz\per\second}$, carried out on the distributed computing  volunteer project \EaH{} and the ATLAS supercomputer at the Max Planck Institute (MPI) for Gravitational Physics in Hannover. The frequency-spin-down range is based on the predictions of \cite{Pagliaro2023}, according to which more than 95\% of the potentially detectable sources lie in this range.

We use the public data of the third observing run (O3) of the two Advanced LIGO detectors, near Hanford (LHO) and Livingston (LLO), respectively \citep{LIGOScientific:2014pky,LIGOScientific:2019lzm}. Since continuous-wave searches are computationally limited, the investment of computing resources in the processing of an additional data stream has to be carefully weighted against the gains in sensitivity from it. Our investigations advise against including O3 data from the Virgo detector in this search, due to its lower sensitivity \citep{Virgo:2014yos,LIGOScientific:2023vdi}.

We use a staged approach: we search half of the data (O3a) and keep the other half (O3b) to verify any candidate that survives the first search. The O3a search is actually a hierarchy of seven stages, beginning with a computationally intensive step, which is also farmed out on the volunteer computing project Einstein@Home. Finally, any candidate surviving the O3b stage (Stage 8) is confirmed with a fully coherent search on the entire data set, O3a\texttt{+}b, based on which the signal parameters are most accurately estimated.

The plan of the paper is as follows: Section \ref{sec:signal} describes the signal model and \ref{sec:SearchGeneral} the search methodology. The \EaH{} search is described in Section \ref{sec:stage0}; the hierarchical follow-ups in Section \ref{sec:followups}. Results are presented in Section \ref{sec:results} and conclusions in \ref{sec:conclusions}.

\section{The signal}\label{sec:signal}

The waveforms $h(t)$ that we target in this search are fairly simple: nearly monochromatic signals with frequency and amplitude modulation due to the Earth's motion.  At the gravitational-wave detector, they take the form \citep{Jaranowski:1998qm}: 
\begin{equation}
	h(t)=F_+ (\alpha,\delta,\psi ;t) h_+ (t) + F_\times (\alpha,\delta,\psi; t) h_\times(t),
	\label{eq:signal}
\end{equation}
where $F_+ (\alpha,\delta,\psi;t)$ and $F_\times(\alpha,\delta,\psi;t)$ are the detector beam pattern functions for the ``+" and ``$\times$" polarizations, $(\alpha,\delta)$ are the right-ascension and declination of the source, $\psi$ is the polarization angle
and $t$ is the time at the detector. The waveforms $h_+ (t)$ and $h_\times (t)$ take the form
\begin{eqnarray}
	h_+ (t)  =  A_+ \cos \Phi(t) \nonumber \\
	h_\times (t)  =  A_\times \sin \Phi(t),
	\label{eq:monochromatic}
\end{eqnarray}
with the ``+" and ``$\times$" amplitudes 
\begin{eqnarray}
	A_+  & = & {1\over 2} h_0 (1+\cos^2\iota) \nonumber \\
	A_\times & = &  h_0  \cos\iota. 
	\label{eq:amplitudes}
\end{eqnarray}
$h_0\geq 0$ is the intrinsic gravitational-wave amplitude, $0\leq \iota \leq \pi$ is the angle between the total angular momentum of the star and the line of sight, and $\Phi(t)$ is the phase of the gravitational-wave signal at the time $t$.
 If $\tau_{\mathrm{SSB}}$ is the arrival time of the wave with phase $\Phi(t)$ at the solar system barycenter, then $\Phi(t)=\Phi(\tau_{\mathrm{SSB}}(t))$. The gravitational-wave phase as a function of $\tau_{\mathrm{SSB}}$ is assumed to be 
\begin{multline}
	\label{eq:phiSSB}
	\Phi(\tau_{\mathrm{SSB}}) = \Phi_0 + 2\pi [ f(\tau_{\mathrm{SSB}}-{\tau_0}_{\mathrm{SSB}})  +
	\\ {1\over 2} \dot{f} (\tau_{\mathrm{SSB}}-{\tau_0}_{\mathrm{SSB}})^2 ].
\end{multline}
We take ${\tau_0}_{\mathrm{SSB}}=$ \paramTref\ (Barycentric Dynamical Time in GPS seconds) as a reference time.

We assume that in our target population, the following quantities are uniformly distributed: $\SI{\paramfmin}{\hertz} \leq f \leq \SI{\paramfmax}{\hertz}$, $\left|\cosi\right| \leq 1$, $\left|\psi\right| \leq \pi/4$, source position $0 \leq \alpha < 2\pi$ and $\left|\sin\delta\right| \leq 1$ each distributed uniformly. We assume that the spin-down is distributed log-uniformly in our search range, reflecting our ignorance of the actual spin-down distribution. 
\par\noindent
As we will see, various parameters pertaining to the multistage search presented in this paper are set based on the recovery rate of the search performed over the same reference signal population. Now we describe this reference population. We use $\injRecoveredApprox$ signals. The frequency, spin-down, position, $\cos\iota$ and $\psi$ parameters are distributed as described above. The amplitudes are such that the sensitivity depth (defined in Eq.~\ref{eq:SD}) is uniformly distributed in ${\cal{D}}\in[50,65]~[1/\sqrt{\textrm{Hz}}]$, bracketing a competitive but realistic sensitivity depth value of $56 ~[1/\sqrt{\textrm{Hz}}]$.

\section{Generalities of the searches}
\label{sec:SearchGeneral}

\subsection{The data}
\label{subsec:data}

\begin{figure}[h!tbp]
	\includegraphics[width=\columnwidth]{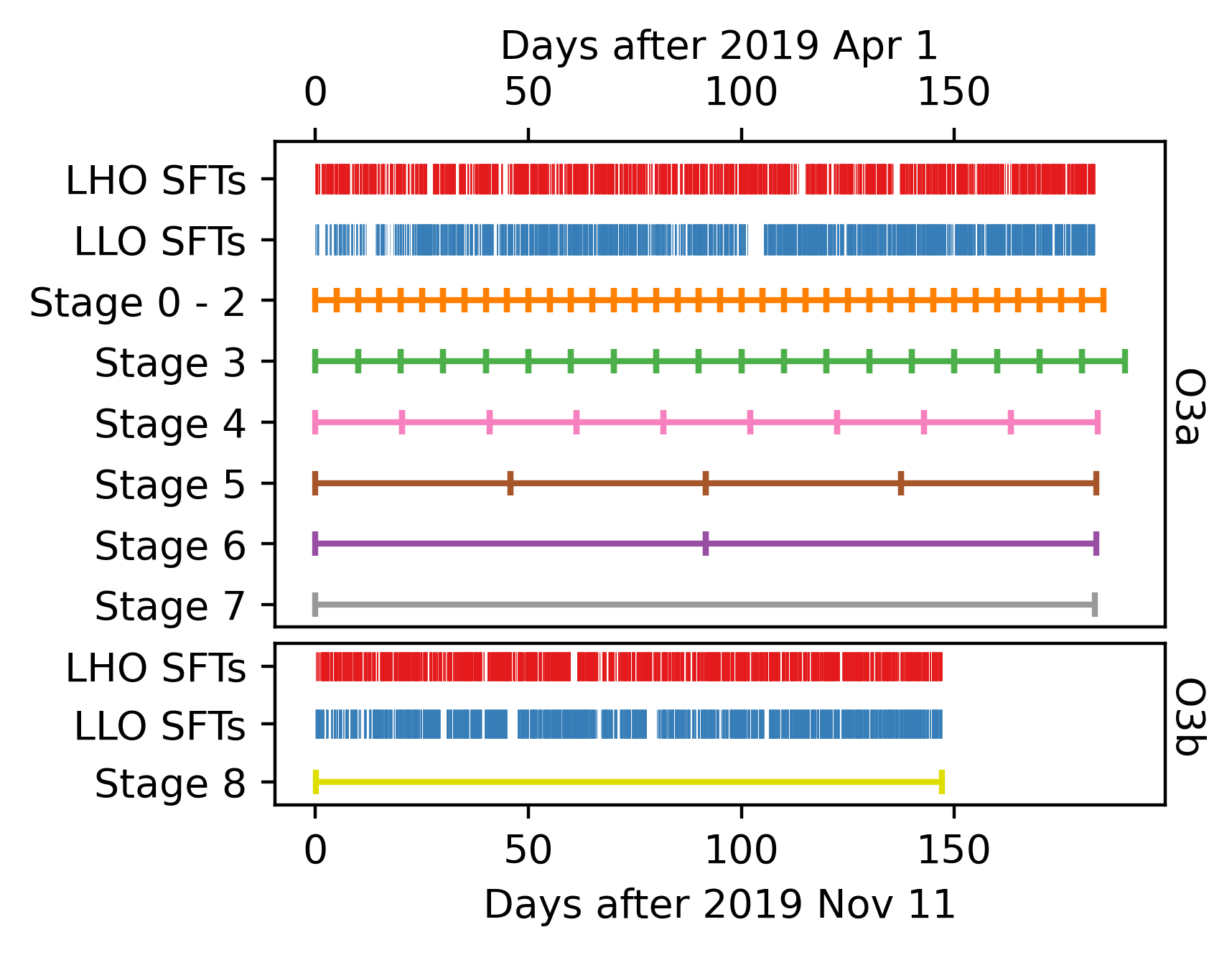}
	\caption{Segmentation for the various searches and input data (SFTs in LHO and LLO). Stages 7, 8, and 9 are fully coherent using O3a, O3b, and O3a\texttt{+}b respectively.} 
	\label{fig:NsegSegmentation}
\end{figure}

We use O3 calibrated data with linear and nonlinear noise subtraction \citep{Davis:2018yrz,Vajente:2019ycy}, which removes spurious noise due to laser beam jitter, calibration lines, power mains, and background noise.

As with previous Einstein@Home searches, we remove noise negatively affecting our search, namely \emph{lines} in the frequency and  \emph{glitches} in the time domain, as described in \citep{Steltner:2021qjy}.

The data are split into our usual format of short time-baseline Fourier transforms (SFTs) with a half-hour time baseline. These are grouped into segments of variable duration for the different coherence times employed in the follow-up, as shown in Figure \ref{fig:NsegSegmentation}.

\subsection{The search}
\label{subsec:search}

We utilize two detection statistics, the $\F$-statistic \citep{Jaranowski:1998qm, Cutler:2005hc} and the line- and transient-line-robust statistic $\bsgltl$ \citep{Keitel:2015ova}. The data are split into $\Nseg$ segments of equal span $\Tcoh$. The data of both detectors in each segment are combined coherently and the detection statistic values for each segment are calculated. The $\F$-statistic values from the $\Nseg$ segments are summed to yield the final semicoherent detection statistic:
\begin{equation}
\hat\F(x,{\bf{\lambda}_t}) = {1\over {\Nseg}} \sum_{i=1}^{\Nseg} \F_i(x,{\bf{\lambda}_t}),
\end{equation}
where $x$ indicates the data and ${\bf{\lambda}_t}$ the parameters defining the template waveform. For an isolated rotating neutron star, the template waveform is defined by the gravitational-wave frequency $f$, its derivative $\dot{f}$ (often spin-down) and the sky position $(\alpha, \delta)$: ${\bf{\lambda}}=({f,\dot{f},\alpha,\delta})$.

The $\hat\F$-statistic is computed from the log-likelihood ratio of the signal hypothesis to the Gaussian noise hypothesis, whereas the $\bsgltl$-statistic tests the signal hypothesis against an expanded noise hypothesis, i.e. ``G'' Gaussian noise or ``L'' lines  or ``tL'' transient lines \citep{Keitel:2015ova}. So, while the $\F$-statistic is susceptible to disturbances generated by spectral lines, the $\bsgltl$-statistic greatly reduces the number of candidates from these disturbances. Therefore, we rank the results using the latter.

For efficiency reasons, the detection statistic value is first computed on a coarse template grid, and then approximated on a finer grid \citep{Pletsch:2010xb}. At the end, the detection statistic of the highest-ranking results is recomputed exactly at the fine-grid template point. The recomputed quantities are indicated by a subscript $r$. These are the results returned to the \EaH{} central server. We refer to the waveform templates and the associated detection statistic values of the returned results as \emph{candidates}. 

\begin{figure}[h!tbp]
    \includegraphics[width=\columnwidth]{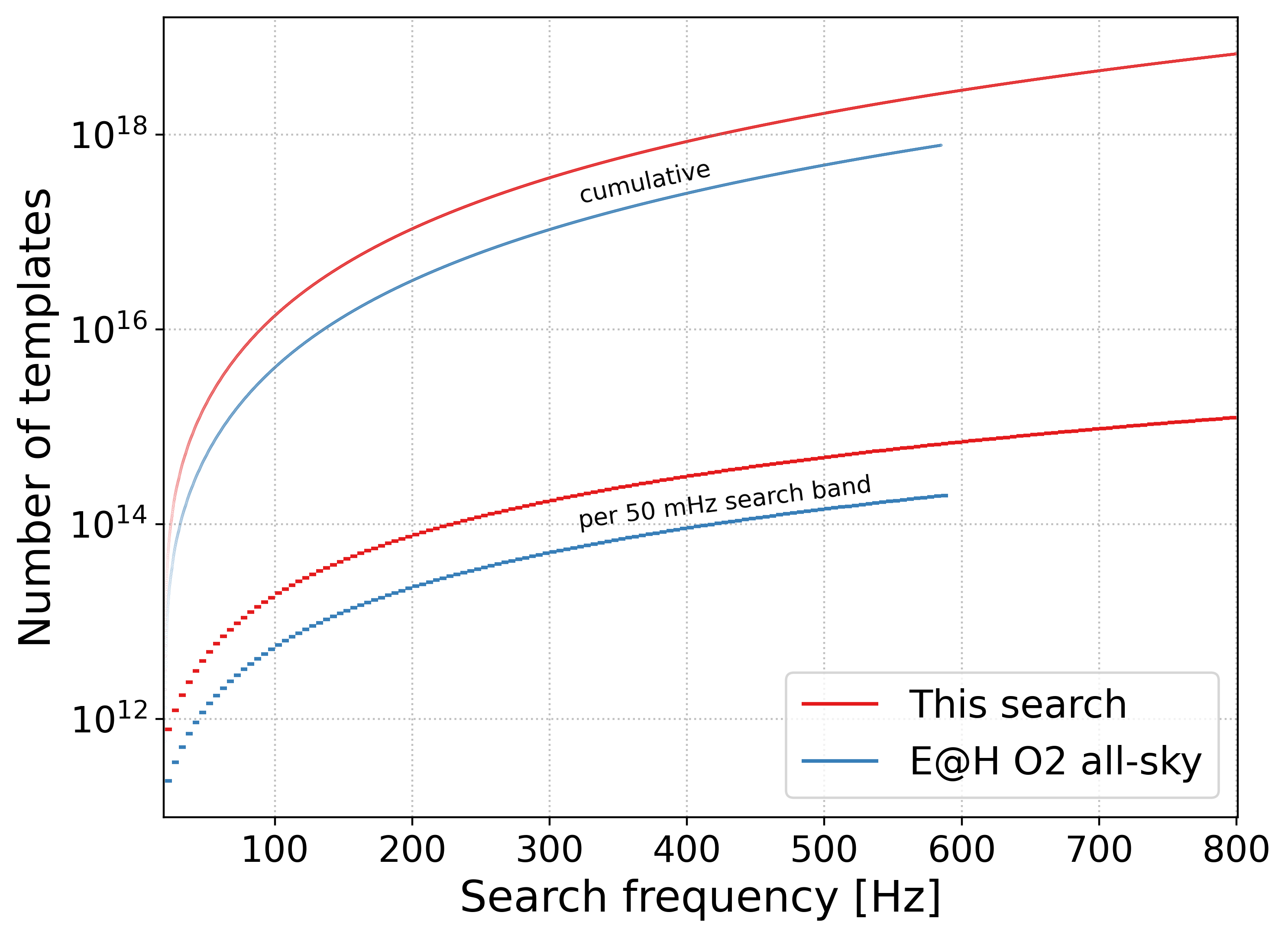}
\caption{Number of templates searched, per $\fiftyMHzband$ band and cumulative, by this search and by the \EaH{} O2 all-sky search \citep{Steltner:2020hfd} ending at $\SI{585.15}{\hertz}$. The sky resolution increases with frequency (see Eq.~\ref{eq:skyGridSpacing}), and so does the number of templates. This search uses a finer resolution than \citet{Steltner:2020hfd} for the same range in spin-down and sky, thus requiring more templates overall.}  
\label{fig:NumberOfTemplatesIn50mHz}
\end{figure}

The grid spacings are chosen to minimize the computational cost for a given average loss in the detection statistic due to signal/template mismatch -- this quantity is known as the ``average mismatch" and is indicated with $\left<\mu\right>$. The average mismatch value is chosen based on computational feasibility. Table \ref{tab:FUtable} shows the spacings and the average mismatch for all stages. The first search -- Stage 0 -- is the most challenging, because the computational cost of surveying the entire parameter space is very high, and this results in a grid with a high mismatch -- about 56\%. Figure~\ref{fig:NumberOfTemplatesIn50mHz} shows the number of templates in this grid, as a function of frequency. The total number of templates searched is $\paramtotaltemplates$. The total number of coarse-grid templates is $\approx 2.7\times 10^{16}$.

The grids in frequency and spin-down are defined by $\delta f$ and $\delta\fdot$, respectively, and these do not change across the search range. Conversely, the sky grid varies with frequency, becoming finer at higher frequencies. Our sky grids are approximately uniform on the celestial sphere orthogonally projected on the ecliptic plane and are defined by the parameter $m_{\text{sky}}$. For the equations defining the projected coordinates, see Eq.s (14) and (15) in \citep{Singh:2017kss}. The tiling is a hexagonal covering of the unit circle each hexagon having the edge length $d$:
\begin{equation}
d(m_{\text{sky}})=0.15 \sqrt{ m_{\text{sky}} } \left [{{100~\textrm{Hz}}\over f}\right].
\label{eq:skyGridSpacing}
\end{equation}

\label{subsec:grids}

\section{Stage 0: the first search}
\label{sec:stage0}

\subsection{The distribution of the computational load}
\label{sub:loadOnEaH}

The first eight searches use LIGO data from the first half of the third observing run (O3a), i.e. between GPS time \dataStartGPS{} (\dataStartDate) and \dataEndGPS{} (\dataEndDate). 

All stages employ the ATLAS cluster. Stage 0 additionally leverages the computing power of the \EaH{} project. \EaH{} is a distributed volunteer computing project built upon the BOINC infrastructure \citep{Boinc2, Boinc3, Boinc1}, where volunteers can spend their idle computational resources to solve scientific problems that require large amounts of computing power. ATLAS is the supercomputer cluster at the MPI for Gravitational Physics in Hannover\footnote{\texttt{https://www.atlas.aei.uni-hannover.de/}}. 

This is the first Einstein@Home continuous gravitational-waves search primarily run on Graphical Processing Units (GPUs). The advantage of using GPUs is that certain instructions can be efficiently parallelized, improving the performance by more than an order of magnitude, compared to CPUs. 

The search is split into work-units (WUs), which run on an average GPU for 10-30 minutes. 
A total of $\paramtotalWUsmillions$ million WUs were computed, totaling over $2000$ years of computing on a single GPU. Each WU searches $\paramWUtotaltemplates$ template waveforms, corresponding to half Hz in frequency, the full spin-down range and a portion of the sky, and returns a so-called ``top-list" containing the top-ranking $\nCandPerWUHalfHz$ results.

A fraction of the top-list results returned to the server are considered for further processing. In general, the more results that are considered, the lower is the smallest detectable signal. For every search, we consider as many results as we possibly can, given computational constraints.

The overall number of top-list results is $\paramNCandTotal$, which effectively is about a factor of $\approx$ 10 higher compared to the previous all-sky Einstein@Home search \citep{Steltner:2020hfd}. This is consistent with the fact that we search more templates compared to \citet{Steltner:2020hfd}, as Fig. 2 shows.

The fraction of the top-list results that we consider for further processing is comparable to our previous search. But since the number of top-list results is $\approx$ 10 times larger than our previous search, we have to process $\approx$ 10 times more results. The first step in processing so many more results is enabled by a new and enormously more efficient clustering method \citep{Steltner:2022aze}.

Despite the significant achievements of the cleaning efforts, there are still disturbances in the data, which, if loud enough, can saturate the entire half-Hz top-list and render it useless. Such disturbances are however typically concentrated in a frequency band much smaller than half Hz, so to avoid them saturating the entire half-Hz top-list, the search code on the volunteer host maintains 10 {\it{independent}} top-lists, one for each of the ten 50 mHz sub-bands in the half-Hz band. 
The final half-Hz top-list is the union of all 10 of the 50 mHz top-lists. Each of 50 mHz top-list comprises $\nCandPerWUfiftymHz$ results.

\begin{figure}[h!tbp] 
	\centering
	\includegraphics[width=\columnwidth]{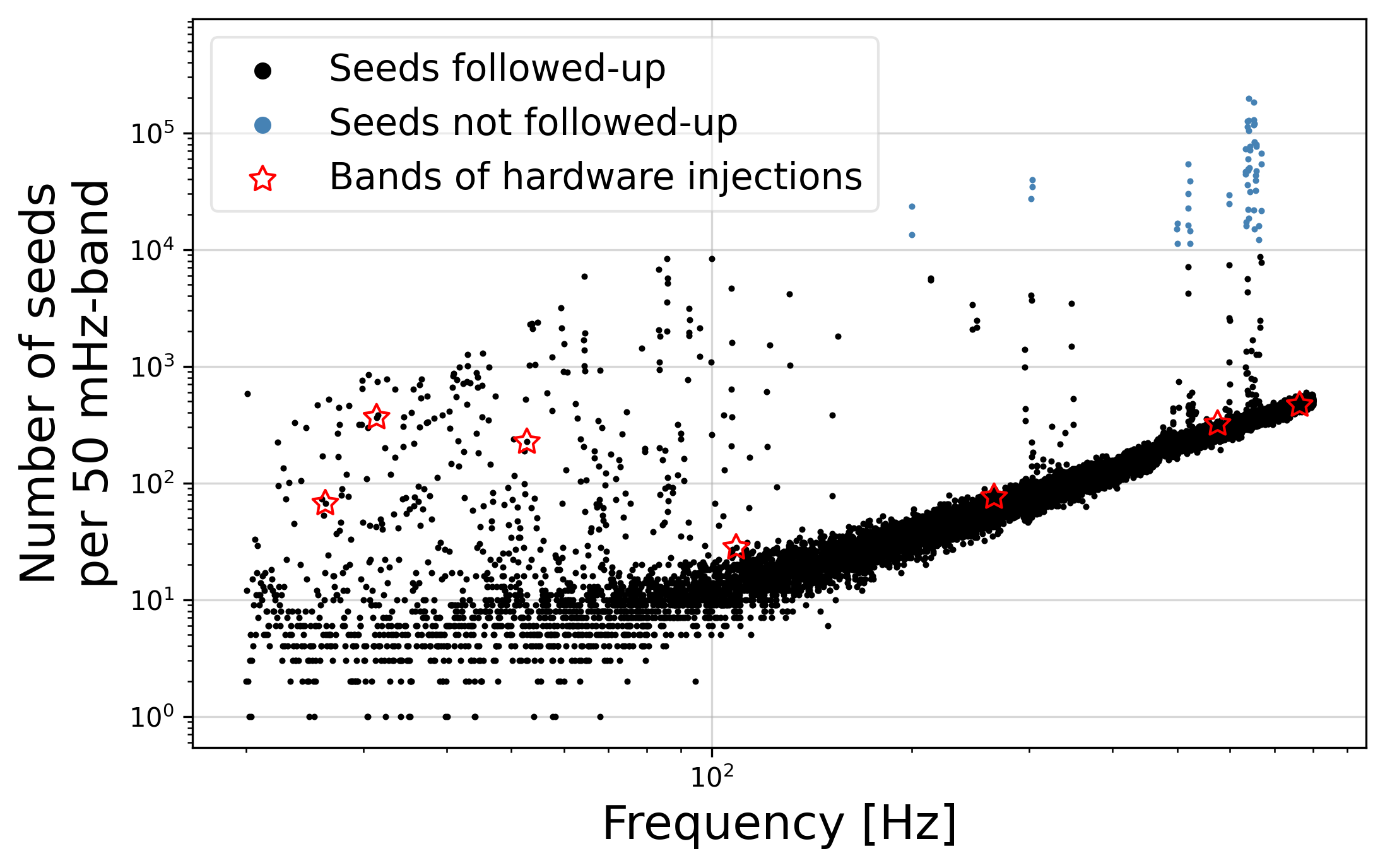}
	\caption{Stage 0 candidates from every \fiftyMHzband{} band. Candidates from bands with more than $\num{10000}$ candidates per band are not followed up; the rest are.}
	\label{fig:seedsPerBand}
\end{figure}
\begin{figure*}[h!tbp] 
	\centering
	\includegraphics[width=\textwidth]{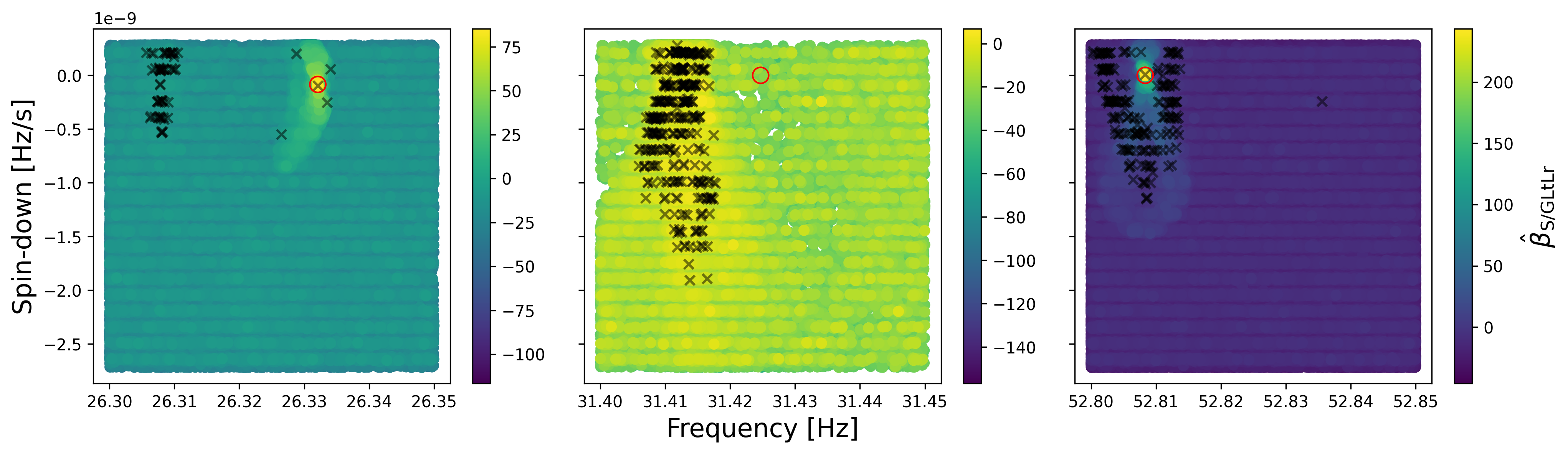}
	\caption{Frequency-spin-down plots for the three low-frequency hardware injections with IDs 10 (left), 11 (middle), and 5 (right; \citet{O3_injection_params}). Color-coded are the detection-statistic of search results within \fiftyMHzband{} of the injection parameters, the full spin-down range and the whole sky. The black crosses mark candidates that are followed up. The red circle indicates the hardware injection parameters. The plots showcase three interesting cases. \emph{Left}: a relatively loud hardware injection is recovered in the vicinity of a disturbance. \emph{Middle}: this shows the only not-recovered injection (ID 11). The larger detection statistic values and candidates in this band have nothing to do with the injection signal, but come from a disturbance. \emph{Right}: the very loud hardware injection (ID 5) also leads to enhanced detection statistic values and candidates at nearby parameter points.}
	\label{fig:threeLowFreqHIs}
\end{figure*}

\subsection{Post-processing}
\label{sec:postProc}

The following post-processing steps are performed on the results returned from the first search:
\begin{description}
	\item[Banding] 
	all results for each $\fiftyMHzband$ band -- from the full spin-down range and all sky points -- are gathered. Based on them, a series of diagnostics are produced, which help identify bands affected by disturbances (see Section IIIB in \citet{LIGOScientific:2017wva}). We find that $\approx$ 1.3\% of the $\fiftyMHzband$ bands are disturbed, but they contribute 20\% of the candidates.  As explained in Section \ref{sec:ULs}, these bands will be excluded from the upper limit statements, but candidates from these bands will in general be followed up.

	\item[Clustering] 
	since our search grids are somewhat oversampled to reduce the signal-to-template waveform mismatch, nearby templates are not independent. Hence, a disturbance or a signal produces some to many nearby results, while statistical fluctuations rarely ``clump". Our clustering method identifies results that are due to the same root cause, so that they can be considered as a single candidate. 
	
	Every cluster is identified by ``a seed", i.e. a set of signal parameters ${\bf{\lambda}}_{seed}$, and by an uncertainty range $\Delta{\bf{\lambda}}$. The signal parameters are different for every cluster, whereas the uncertainty range is exactly the same for all clusters. The meaning of the uncertainty range $\Delta{\bf{\lambda}}$ is the following: $> 99.9\%$ of signals of the reference population described at the end of Section~\ref{sec:signal} give rise, after Stage 0, to at least a cluster whose seed parameters are within a distance $\Delta{\bf{\lambda}}$ of the signal parameters.
	
	We indicate the uncertainty intervals with $\Delta f$, $\Delta\fdot$ and $r_\textrm{sky}$. The uncertainty region in the sky is a circle in the orthogonally projected ecliptic plane centered at the candidates' sky position, with radius $r_\textrm{sky}$.
	
	\par\noindent
The clustering parameters are determined based on search-and-recoveries on the reference signal population after Stage 0. A clustering setup is chosen that minimizes the amplitude of the weakest signals recovered, while the number of false alarms remains below a given threshold, determined by the total amount of time we want to devote to the Stage 1 follow-up. With a Stage 1 follow-up of a few weeks, the 90\% recovery rate of the chosen clustering setup corresponds to a population of signals with amplitude $h_0$ such that $\Depth = 56 ~[{1/\sqrt{\textrm{Hz}}}]$. For a fake signal to be counted as recovered, there needs to exist a seed which can be associated with the injection, that was not there in the data without the injection.

We remind the reader that the sensitivity depth $\Depth$, first introduced by \cite{Behnke:2014tma}, is defined as 
\begin{equation}
\Depth = {\sqrt{S_n(f)}}/{h_0(f)}~[{1/\sqrt{\textrm{Hz}}}],
\label{eq:SD}
\end{equation} 
with $h_0(f)$ being the continuous gravitational-wave intrinsic strain amplitude. If $h_0$ is the upper limit from a search at frequency $f$, then $\Depth$ describes the sensitivity of that search in terms of ``how deep" below the noise level $S_n$ the search can detect signals. But Eq.~\ref{eq:SD} can also be seen as defining the amplitudes of the population of signals at different frequencies, which would be $\approx$ equally well detected by a given search pipeline: fix the detection pipeline, this determines the value of $\Depth(f)$, and Eq.~\ref{eq:SD} gives the amplitude $h_0(f)$ of the smallest detectable signal. This is used in search-and-recovery simulations that aim at characterizing the detection efficiency of a pipeline, or a piece thereof, as done above for the clustering.
	
	\par\noindent
	The clustering reduces the $\paramNCandTotal$ results to a more manageable set of $\approx 3.5$ million seeds. For the remainder of the paper, we may also refer to the cluster seeds as the ``Stage 0 candidates" or simply as ``candidates". 
	
	\item[Follow-Up] we follow-up the 3.5 million candidates as detailed in the next section. The average number of candidates per 50 mHz band ranges between $\approx$ 50-350 candidates, as shown in Figure \ref{fig:seedsPerBand}.
	We do not follow up any candidate from 50 mHz bands with more than $\num{10000}$ candidates, as this is a clear indication that the band is affected by disturbances and the candidates in it are due to the disturbances. \noFUBands{} bands are hence excluded from the follow-up and they are listed in the Supplemental Materials and at \citep{O3AS-AEI}. These bands are also excluded from the upper limit statements.

	\par\noindent 
	From Figure \ref{fig:seedsPerBand} we see that four of the seven fake signals added to the data for validation -- the so-called hardware injection signals -- are in relatively ``quiet" bands (the ones at higher frequencies), and three (at $\sim 26.3, 31.4$, and $52.8$ Hz) are instead in bands that are clearly affected by some excess. The 52.8 Hz injection is very loud and it is solely responsible for the excess. The 31.4 Hz injection produces a very weak signal in the search results -- in fact, this is the injection that we are not able to detect -- and the excess is due to a disturbance. The 26.3 Hz injection is detected, but the excess comes from a disturbance also present in the band. Figure \ref{fig:threeLowFreqHIs} shows the search results in these three bands and illustrates these three different situations. More information on the hardware injections is given in Section \ref{sec:HIs}.
		\end{description}

\begin{figure*}[h!tbp] 
	\centering
	\includegraphics[width=\textwidth]{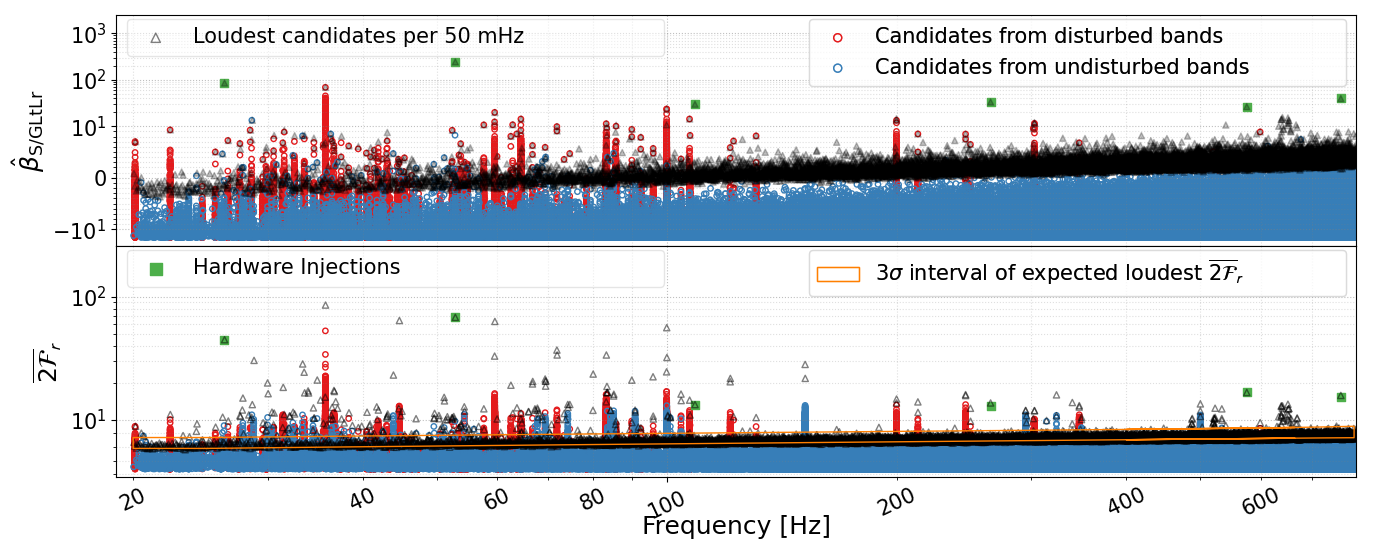}
	\caption{Detection statistic value in $2\avF_r$ and $\bsgltlr$ of the loudest candidate in each \fiftyMHzband{} (gray triangles, which appear black when there is many of them superimposed upon one another) and of each candidate selected for follow-up (circles). This shows that the candidates that we are following up have $2\avF_r$ values up to $40\%$ smaller than if only the loudest per 50 mHz band had been selected. The hardware injections are denoted in green. The $3\sigma$ interval around the expected highest $2\avF_r$ determined by the number of searched templates, is given in orange and is in good agreement with our data. We also mark (red circles) follow-up candidates from disturbed bands (see bullet-point ``Banding" in Section~\ref{sec:postProc}).}
	\label{fig:loudestHalfHzAndFU0CandidatesVersusFreq}
\end{figure*}

An overview of all Stage 0 search results is given in Figure~\ref{fig:loudestHalfHzAndFU0CandidatesVersusFreq}.

\section{The follow-up searches}
\label{sec:followups}

\begin{deluxetable*}{lcccccccccccccccc}
	\tablecaption{Overview of the full hierarchy of searches. Shown are the values of the following parameters: the stage number, the coherent time baseline $T_\mathrm{coh}$ of each segment and the number of segments $\Nseg$; the grid spacings $\delta f,\delta\fdot$ and $m_{\text{sky}}$; the average mismatch $\left<\mu \right>$; the parameter space volume searched around each candidate, $\Delta f, \Delta\fdot$ and ${r_\textrm{sky}}$. The radius  ${r_\textrm{sky}}$ is expressed in units of the side of the hexagon sky-grid tile of the Stage 0 search (Eq.~\ref{eq:skyGridSpacing});
	the number of candidates searched (N$_{\textrm{in}}$ ) and how many of those survive and make it to the next stage (N$_{\textrm{out}}$). \label{tab:FUtable}	
	}
	\tablehead{
		\colhead{Search} & \colhead{$T_\mathrm{coh}$} & \colhead{$\Nseg$} & \colhead{$\delta f$} & \colhead{$\delta {\dot{f}}$} &  \colhead{$m_{\text{sky}}$} & \colhead{$\left<\mu \right>$ } & \colhead{$\Delta f$} & \colhead{$\Delta\fdot$} & \colhead{$ {r_\textrm{sky}\over {d(\mskyFUZero )}}$} &  \colhead{ N$_{\textrm{in}}$ }& \colhead{N$_{\textrm{out}}$} \\
		& hr &  & $\mu{\textrm{Hz}}$ &  {{ $10^{\scriptstyle{-14}}$} Hz/s} &  & $10^{-2}$ & $\mu{\textrm{Hz}}$ & {{ $10^{\scriptstyle{-14}}$} Hz/s} &  & & 
	}
	\startdata
\TBstrut Stage 0 & $\TcohFUZero$ 	& $\NSegFUZero$ 	& $\dfreqmuHzFUZero$ 	& $\dfdotfgTenfHzFUZero$ 	&   $\mskyFUZero$ & $\meanMismatchFUZero$ & \tiny{full range} 	& \tiny{full range}	& \tiny{all-sky}  & $\paramtotaltemplates$  & $\NCandFUOne$ \\
\TBstrut Stage 1 & $\TcohFUOne$ 	& $\NSegFUOne$ 	&  $\dfreqmuHzFUOne$ 	& $\dfdotfgTenfHzFUOne$	&  $\mskyFUOne$  & $\meanMismatchFUOne$ & $\SRfreqmuHzFUOne$ 	& $\SRfdotFUOne$ 	& $\SRskyFUOne$  &   $\NCandFUOne$ & \NCandFUTwo  \\
\TBstrut Stage 2 & $\TcohFUTwo$ 	& $\NSegFUTwo$ 	&  $\dfreqmuHzFUTwo$ 	& $\dfdotfgTenfHzFUTwo$	&  $\mskyFUTwo$  & $\meanMismatchFUTwo$ & $\SRfreqmuHzFUTwo$ 	& $\SRfdotFUTwo$ 	& $\SRskyFUTwo$ 	&  $\NCandFUTwo$ & \NCandFUThree  \\
\TBstrut Stage 3 & $\TcohFUThree$ 	& $\NSegFUThree$ 	&  $\dfreqmuHzFUThree$ 	& $\dfdotfgTenfHzFUThree$	&  $\mskyFUThree$  & $\meanMismatchFUThree$ & $\SRfreqmuHzFUThree$ 	& $\SRfdotFUThree$ 	& $\SRskyFUThree$  &   $\NCandFUThree$ & \NCandFUFour  \\
\TBstrut Stage 4 & $\TcohFUFour$ 	& $\NSegFUFour$ 	&  $\dfreqmuHzFUFour$ 	& $\dfdotfgTenfHzFUFour$&  $\mskyFUFour$  & $\meanMismatchFUFour$ & $\SRfreqmuHzFUFour$ 	& $\SRfdotFUFour$ 	& $\SRskyFUFour$  &   $\NCandFUFour$ & \NCandFUFive  \\
\TBstrut Stage 5 & $\TcohFUFive$ 	& $\NSegFUFive$ 	&  $\dfreqmuHzFUFive$ 	& $\dfdotfgTenfHzFUFive$&  $\mskyFUFive$  & $\meanMismatchFUFive$ & $\SRfreqmuHzFUFive$ 	& $\SRfdotFUFive$ 	& $\SRskyFUFive$  &   $\NCandFUFive$ & \NCandFUSix  \\
\TBstrut Stage 6 & $\TcohFUSix$ 	& $\NSegFUSix$ 	&  $\dfreqmuHzFUSix$ 	& $\dfdotfgTenfHzFUSix$	&  $\mskyFUSix$  & $\meanMismatchFUSix$ & $\SRfreqmuHzFUSix$ 	& $\SRfdotFUSix$ 	& $\SRskyFUSix$ 	&   $\NCandFUSix$ & \NCandFUSeven  \\
\TBstrut Stage 7 & $\TcohFUSeven$ 	& $\NSegFUSeven$ 	&  $\dfreqmuHzFUSeven$ 	& $\dfdotfgTenfHzFUSeven$	&  $\mskyFUSeven$  & $\meanMismatchFUSeven$ & $\SRfreqmuHzFUSeven$ 	& $\SRfdotFUSeven$ 	& $\SRskyFUSeven$ 	 &   $\NCandFUSeven$ & \NCandFUEight  \\
\TBstrut Stage 8 & O3b coh. 	& $\NSegFUEight$ 	&  $\dfreqmuHzFUEight$ 	& $\dfdotfgTenfHzFUEight$	&  $\mskyFUEight$  & $\meanMismatchFUEight$ & $\geq\SRfreqmuHzFUEight$\tablenotemark{a}	& $\SRfdotFUEight$ 	& $\SRskyFUEight$ &   $\NCandFUEight$ & \NCandFUNine  \\
\TBstrut Stage 9\tablenotemark{b} & O3a\texttt{+}b coh.	& $\NSegFUNine$ 	&  $\dfreqmuHzFUNine$ 	& $\dfdotfgTenfHzFUNine$	&  $\mskyFUNine$  & $\meanMismatchFUNine$ & $\geq\SRfreqmuHzFUNine$\tablenotemark{a} 	& $\SRfdotFUNine$ 	& $\SRskyFUNine$  &   $\NCandFUNine$ & \NCandFUNine \\
	\enddata
	\tablenotetext{a}{Since the reference time of the O3b search is different than the reference time of the O3a stages, an uncertainty in spin-down value produces an uncertainty in frequency. The nominal $\SRfreqmuHzFUEight~\mu$Hz value only holds if the signal spin-down were precisely known.	} 
	\tablenotetext{b}{Since already at the previous stage the only surviving candidates are the hardware injections, 
we carry out this stage to demonstrate the accuracy in signal recovery.}
\end{deluxetable*}

\begin{figure}[h!tbp] 
\includegraphics[width=\columnwidth]{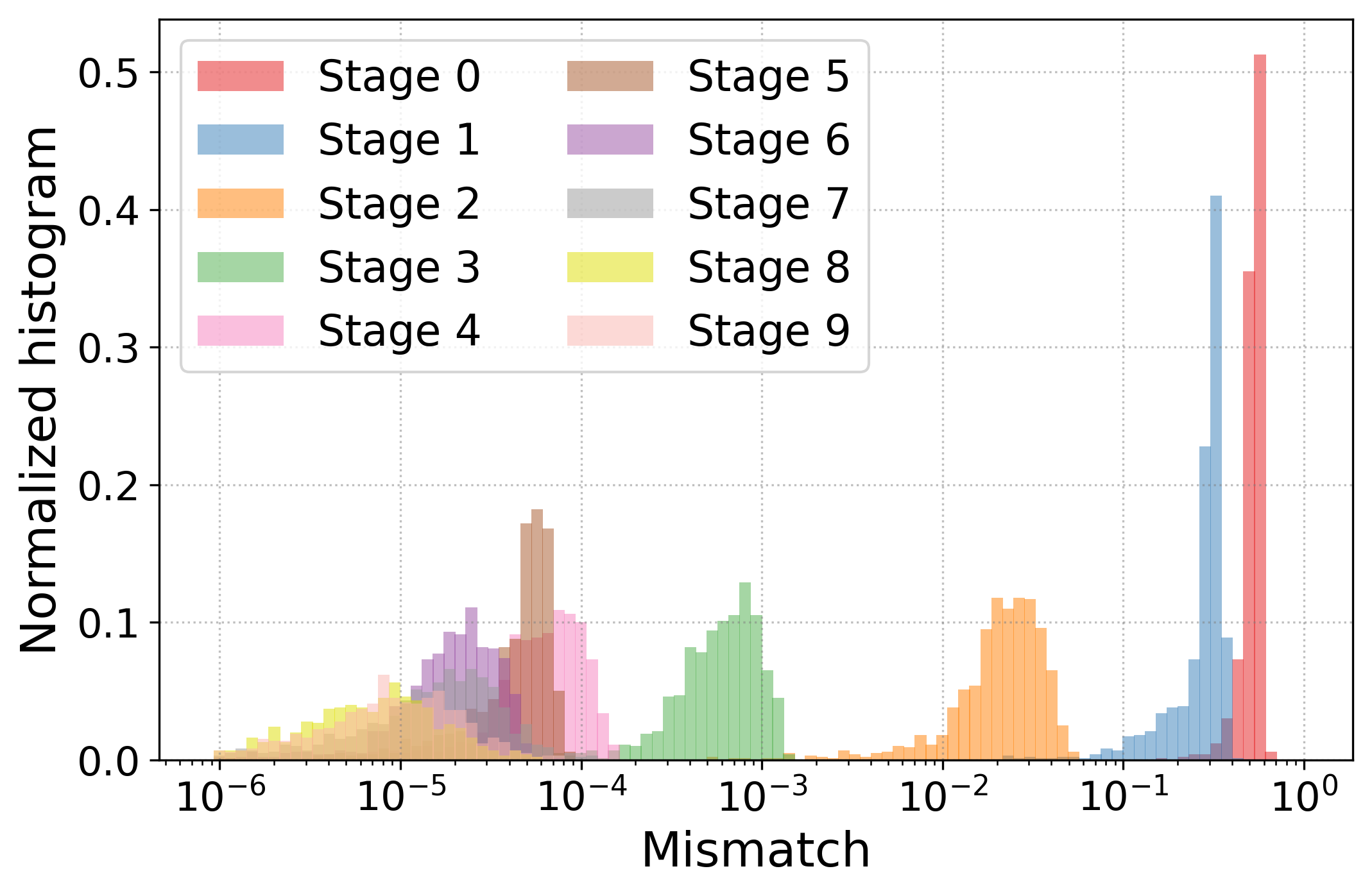}
        \caption{Mismatch distributions of all stages based on 1000 injection-and-recovery Monte Carlo simulations. Due to the high cost of Stage 0, it features a large mismatch of $\sim\meanMismatchFUZero$\%. The setups of later stages are chosen so that the detection statistics increase from one stage to the next, and this is achieved by increasing $\Tcoh$ and/or decreasing the mismatch. The finer setups become possible by the shrinking of the uncertainty region and by the decreasing number of candidates per stage.}
\label{fig:mismatches}
\end{figure}

\begin{deluxetable}{lccc}
	\tablecaption{The threshold value $R^{a}$ used to veto candidates of Stage $a > 3$ (Eq.~\ref{eq:R}); the absolute thresholds used in Stages 1-3: $\TwoFrThreshold$, $\BSGLtLrThreshold$\label{tab:thresholdstable}
	}
	\tablehead{
		\colhead{Search} & \colhead{$R^{a{\textrm{-thr}}}$} & \colhead{$\TwoFrThreshold$} & \colhead{$\BSGLtLrThreshold$} \\
	}
	\startdata
	\TBstrut Stage 1 & $\rVetoFUOne$ & $\aTwoFrFUOne$ & $\aBSGLtLrFUOne$   \\
	\TBstrut Stage 2 & $\rVetoFUTwo$ & $\aTwoFrFUTwo$ & $\aBSGLtLrFUTwo$ \\
	\TBstrut Stage 3 & $\rVetoFUThree$ 	& $\aTwoFrFUThree$ & $\aBSGLtLrFUThree$  \\
	\TBstrut Stage 4 & $\rVetoFUFour$ 	& $\aTwoFrFUFour$ & $\aBSGLtLrFUFour$  \\
	\TBstrut Stage 5 & $\rVetoFUFive$ 	& $\aTwoFrFUFive$ & $\aBSGLtLrFUFive$  \\
	\TBstrut Stage 6 & $\rVetoFUSix$ & $\aTwoFrFUSix$ & $\aBSGLtLrFUSix$  \\
	\TBstrut Stage 7 & $\rVetoFUSeven$ 	& $\aTwoFrFUSeven$ & $\aBSGLtLrFUSeven$  \\
	\TBstrut Stage 8 & $\rVetoFUEight$ 	& $\aTwoFrFUEight$ & $\aBSGLtLrFUEight$ \\
	\enddata
\end{deluxetable}

\begin{figure}[h!tbp]
	\includegraphics[width=\columnwidth]{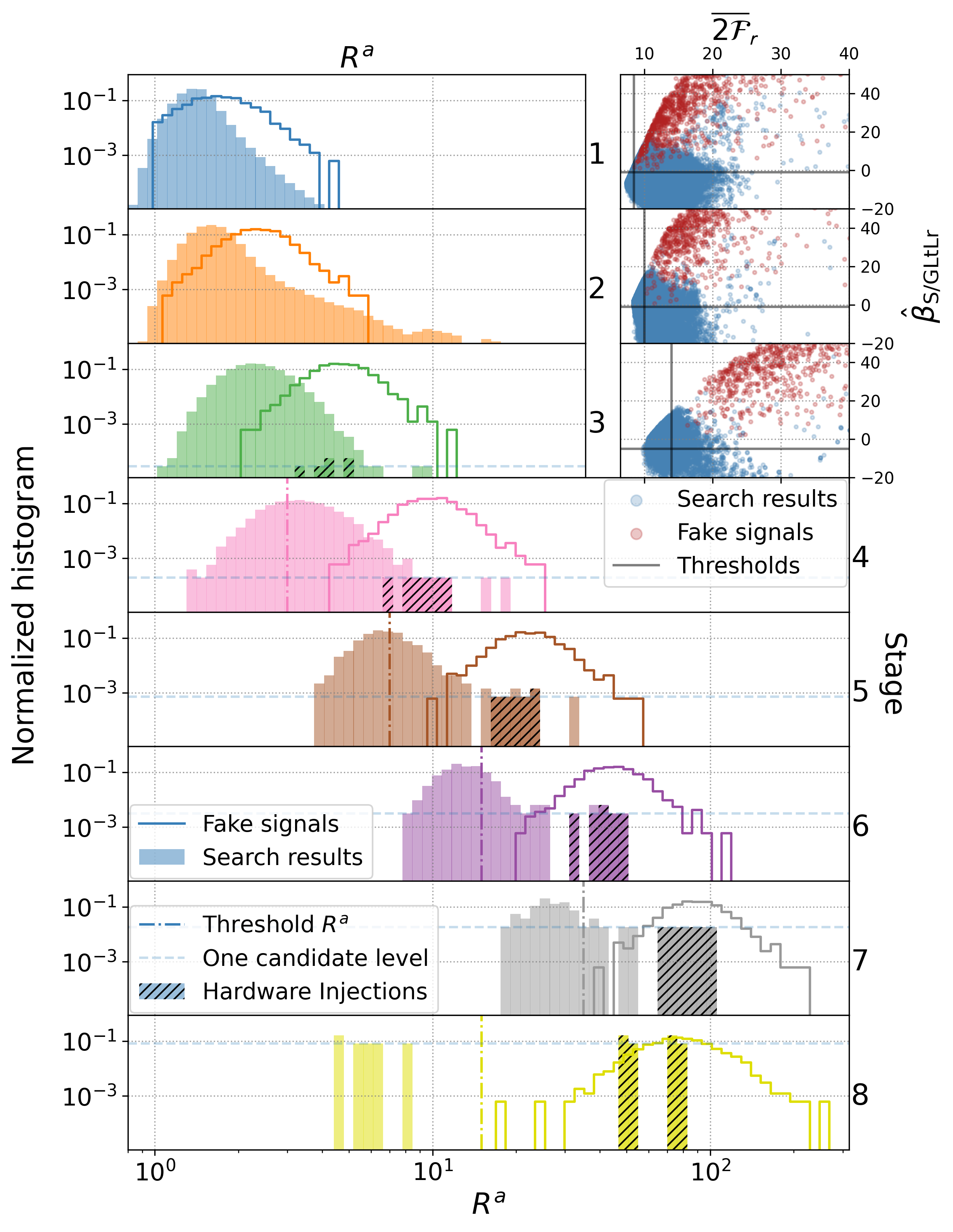}
	\caption{The results of each follow-up stage. The top three right-hand-side panels show a scatter plot of the $(2\avF_r,\bsgltlr)$ values of candidates from the search and candidates from Monte Carlo simulations containing fake signals. The region below the black lines is the candidate rejection region. The wide panels show the distribution of $R^{a}$ for search candidates and for candidates from Monte Carlo simulations containing fake signals. The dashed vertical lines indicate the value of $R^{a{\textrm{-thr}}}$. The dashed horizontal line marks the level where one candidate appears in the normalized histogram.}
	\label{fig:RVeto}
\end{figure}

A hierarchical follow-up of the clustered candidates from Stage 0 is performed. 

An uncertainty region can be defined for each stage, in the same way as done for clustering after Stage 0 (see bullet point ``Clustering" in Section~\ref{sec:postProc}).

At each Stage $n$ the uncertainty region from Stage $n-1$ around each surviving candidate is searched.

From one stage to the next, the grid resolution is increased. For Stages 3-7 the coherence time $\Tcoh$ is $\approx$ doubled, whereas for Stages 1-2 it stays the same as for Stage 0. The reason is that we would not have enough computing power to usefully follow up all Stage 0-1 candidates with a search having a $\Tcoh > 120 $ hrs: the parameter space regions searched in the early stages are much larger than those searched in the later stages and the computational cost per unit volume of parameter space steeply increases with the coherence time $\Tcoh$ \citep{Brady:1998nj}. So in the first stages, we keep the computational cost at bay by not increasing $\Tcoh$. We are however still able to increase the sensitivity of the search by decreasing the grid mismatch, i.e. by using a finer grid. This in turn decreases the uncertainty on the candidate parameters, shrinking the search volume of the next stage. Starting with Stage 3, the search volume is small enough that it becomes computationally feasible to $\approx$ double the coherence time $\Tcoh$ in each stage.

Stages 7, 8 and 9 are fully coherent on O3a, O3b, and O3a\texttt{+}b data, respectively. The search setups and covered regions are the same, as explained in Section~\ref{sec:O3b}.

The parameters defining the search setups and searched/surviving candidates are all given in Table \ref{tab:FUtable}. The mismatch $\mu$ distributions are shown in Figure \ref{fig:mismatches}.

Only the highest detection statistic result from each follow-up is considered, and that becomes the new representative candidate for that stage. 
For stages $a=1,2,3$, a candidate is vetoed unless both its $2\avF_r$ and $\bsgltl{}_r$ values exceed the thresholds given in Table \ref{tab:thresholdstable}. For stages $a=4,\cdots,8$ candidates are vetoed unless their 
\begin{equation}
	R^{a}={2\avF_r^{\textrm{~Stage}~a} - 2\avF_r^{\textrm{~Stage 0}} \over {2\avF_r^{\textrm{~Stage 0}} -4}},
	\label{eq:R}
\end{equation}
exceeds the $R^{a{\textrm{-thr}}}$ threshold value given in Table \ref{tab:thresholdstable}.

The $\TwoFrThreshold$, $\BSGLtLrThreshold$ and $R^{a{\textrm{-thr}}}$ are determined by adding fake signals from our target population (see Section~\ref{sec:signal}) and searching for them, exactly as done in Stage 0 and in the following stages. The total number of fake signals is $\injRecoveredApprox$, and the thresholds are set so that none of the signals are discarded by the vetoes, corresponding to a false dismissal of $< 99.9\%$. The results are shown in the top right-hand-side plots of Figure \ref{fig:RVeto}. 

Of the $3.5$ million candidates that are fed to the hierarchical follow-up, only 12 remain after Stage 7. 

\subsection{Follow-up of the 12 remaining candidates}
\label{sec:O3b}

We follow up the 12 candidates that survive the seven follow-up stages utilizing a different data set, namely the second half of O3 (O3b). Our data span $147$ days, starting at GPS time $1256655667$ (2019 November 1 15:00:49 GMT) and ending at GPS time $1269363493$ (2020 March 27 16:57:55 GMT). We perform a fully coherent search with the same grid spacings as the fully coherent search on O3a data. Since O3b spans a period of time about $\approx 36$ days shorter than O3a, this results in a smaller average mismatch. However, due to the smaller amount of data and to a slightly worse detector sensitivity, all in all, the search sensitivity using O3b data is reduced with respect to O3a by about 15\%. This can be seen in the two bottom panels of Figure~\ref{fig:RVeto}, comparing the respective signal results, and it yields a lower value of $R^{8{\textrm{-thr}}}$.

The uncertainty ranges around the candidates surviving Stages 7 and 8, are larger (by 10\%-30\%) than the uncertainty regions at Stage 6. We hence center the Stage 8 and 9 follow-ups around the corresponding Stage 6 candidates and use the Stage 6 uncertainty region. The Stage 8 and 9 Monte Carlo simulations are of course performed consistently, i.e. based on the Stage 6 candidates surviving Stage 7. The increase in uncertainty for Stages 7 and 8 is somewhat unexpected, and even though we were able to verify it in a number of ways, we do not fully understand its origin.

After the follow-up on O3b data (Stage 8), only six candidates survive, and they are all due to hardware injections.

\section{Results}
\label{sec:results}

\subsection{Recovery of the hardware injections}
\label{sec:HIs}

\begin{deluxetable*}{lccccccc}
	\tablecaption{Frequency, frequency derivative and sky position of the hardware injections and the distance between these and the values recovered by our coherent search using the entire O3 data (Stage 9). The frequency $f$ is given at the reference time 1253764756.0 (GPS time).  \label{tab:HIRecovery}}
	\tablehead{
		\colhead{ID$_{inj}$} & \colhead{ $f$} & \colhead{$\dot{f}$} & \colhead{$\alpha$} & \colhead{$\delta$} & \colhead{$\Delta f$} & \colhead{$\Delta \dot{f}$} & \colhead{Sky distance} \\
		& [Hz] & [Hz/s] & [hr:m:s] & [deg:m:s] & [Hz] & [Hz/s] & [deg:m:s]
	}
	\startdata
	\TBstrut 0 & $\num{ 265.57505348 }$ & $\num{ -4.15e-12 }$ & 4:46:12.4628 & -57:46:57.0510  & $\num{ -4.7e-11  }$ & $\num{ 9.5e-16 }$ &  0:0:0.0741   \\ 
\TBstrut 2 & $\num{ 575.16350527 }$ & $\num{ -1.37e-13 }$ & 14:21:1.4800 & 3:26:38.3626  & $\num{ -1.1e-09  }$ & $\num{ -8.8e-16 }$ &  0:0:0.0955   \\ 
\TBstrut 3 & $\num{ 108.85715939 }$ & $\num{ -1.46e-17 }$ & 11:53:29.4178 & -34:33:48.2313  & $\num{ 6.7e-10  }$ & $\num{ -5.8e-16 }$ &  0:0:0.3080   \\ 
\TBstrut 5 & $\num{ 52.80832436 }$ & $\num{ -4.03e-18 }$ & 20:10:30.3939 & -84:9:39.0964  & $\num{ -6.2e-10  }$ & $\num{ -4.2e-16 }$ &  0:0:0.2212   \\ 
\TBstrut 9 & $\num{ 763.84731649 }$ & $\num{ -1.45e-17 }$ & 13:15:32.5397 & 75:41:22.5205  & $\num{ 9.4e-10  }$ & $\num{ -5.6e-17 }$ &  0:0:0.0023   \\ 
\TBstrut 10 & $\num{ 26.33209638 }$ & $\num{ -8.50e-11 }$ & 14:46:13.3549 & 42:52:38.2953  & $\num{ -8.3e-11  }$ & $\num{ 2.4e-16 }$ &  0:0:0.3109   \\  \\[-1.55em]
	\enddata
\end{deluxetable*}

The hardware injections are signals added to the data by directly moving the detector mirrors in order to provide a check of the entire detection chain. These are de facto reference signals that serve as standard detection benchmarks for any continuous-wave search pipeline.

Seven hardware injections fall in our search range, specifically those with IDs $0, 2, 3, 5, 9, 10$, and $11$ \citep{O3_injection_params}. We recover all but one. 

The missed hardware injection has ID=11, and it is at $\sim$ 31.4 Hz. Its strain amplitude lies just below our upper limit, but its inclination is not particularly unfavorable. The reason why it is not detected is that its parameter values lie at a high mismatch point within the grid, and the resulting detection statistic value is low enough that in the original Stage 0 results there is no candidate associated with this hardware injection.

The O3a\texttt{+}b candidates associated with the hardware injections are within $\pm 10^{-9}$ Hz, $\pm 10^{-15}$ Hz s${}^{-1}$, and $\SI{0.3}{\arcsecond}$ in sky (see Table \ref{tab:HIRecovery}) of the correct parameter values. This remarkable accuracy is expected from long-baseline observations, and it is one of the promises of the science of continuous gravitational waves.

The O3a\texttt{+}b search setup is not optimal; rather, it is a practical search with improved sensitivity with respect to all the previous stages. This means that probably even higher parameter estimation accuracy could be obtained.

Three of the candidates surviving Stage 7 are ``secondaries" associated to hardware injections 2, 9, and 10. Compared to the primaries, they are not as significant and lie at a much greater distance from the true signal values: $> 2\times 10^{-4}$ Hz in frequency and $> 6\times 10^{-11}\,\mathrm{Hz\,s}^{-1}$ in spin-down.  None of the secondaries survive Stage 8. 
\subsection{Upper limits}
\label{sec:ULs}

Based on our null results, we set 90\% confidence frequentist upper limits on the detectable intrinsic gravitational-wave amplitude $h_0$ at the detector. This is the smallest continuous gravitational-wave amplitude such that we can recover 90\% of the signals of our target population. We estimate the upper limits in half-Hz bands with injection-and-recovery Monte Carlo simulations and show the results in Figure \ref{fig:h0ULs}. These are given in machine-readable format in the Supplemental Material and at \citep{O3AS-AEI}.

We employ the same method as in \citet{Steltner:2020hfd}. In a first pass, we add fake signals drawn from the target population, treat the resulting data exactly as is done in the data preparation for the search, search with the Stage 0 setup, and cluster the results as done for the Stage 0 results. In a second pass, we do not add the fake signal, and perform every step exactly alike. We consider the signal to be detected if the results with the added fake signal (1) produce a candidate with parameters close enough (within the uncertainty region) to the signal parameters, and (2) the detection statistic value of that candidate is larger than the detection statistic measured at the same parameter space point in the absence of a signal. We do this in every half-Hz band $200$ times for each $h_0$ value. The confidence $C(h_0)$ is the fraction of detected signals at that $h_0$ value. We consider at least five different $h_0$ values. We fit the resulting $C(h_0)$ curve and derive the upper limit value $h_0^{\textrm{UL}}$ such that $C(h_0^{\textrm{UL}})=0.9$, as described in Section 5.1 of \citet{Fesik:2020tvn}.

Our upper limits do not hold in the \fiftyMHzband\, 
bands that we marked as disturbed. As already said, the disturbed band list is provided in machine-readable format in the Supplemental Material and at \citep{O3AS-AEI}. 
Upper limits are also not given in \halfHzband\, bands where (1) all \fiftyMHzband\, bands are marked disturbed or (2) where the $90\%$ detection efficiency is not reached, due to the line-cleaning procedure removing the added fake signal, as it would have done with a real signal. We find that the upper limit value is impacted by the cleaning in less than 6\% of the half-Hz bands.

Based on our upper limits, we achieve sensitivity depths in the range of  $\left(52-59\right)1/\sqrt{\mathrm{Hz}}$. This constitutes a sensitivity improvement of $\sim 6\%$ stemming solely from the search method, with respect to the previous O2 \EaH{} search \citep{Steltner:2020hfd}. 

The upper limits on $h_0$ can be translated in upper limits on the ellipticity $\varepsilon$ of a source modeled as a triaxial ellipsoid spinning around a principal moment of inertia axis ${I}$ at a distance $d$:
\begin{equation}
	\begin{split}
		\varepsilon^{\textrm{UL}} &=1.4 \times 10^{-6} ~\left( {h_0^{\textrm{UL}}\over{1.4\times10^{-25}}}\right ) \times \\
		&\left ( {d\over{1~\textrm{kpc}}}\right ) \left ({{\textrm{170~Hz}}\over f} \right )^2 \left ({10^{38}~{\textrm{kg m}}^2\over I} \right ).\\
	\end{split}
	\label{eq:epsilon}
\end{equation}
Figure \ref{fig:epsilonULs} shows the upper limits on the ellipticity $\varepsilon$ for different distances.

Another possible emission mechanism is $r$-modes, unstable toroidal fluid oscillations driven by the Coriolis force, emitting at $\approx 4/3$ of the spin frequency \citep{Andersson:1997xt,Friedman:1997uh,Lindblom:1998wf}. We translate the upper limits on the gravitational-wave amplitude $h_0$ into upper limits on the $r$-mode amplitude \citep{Owen:2010ng}:
\begin{equation}
	\alpha^{\textrm{UL}} = 0.028
	\left( \frac{h_0^{\textrm{UL}}}{\num{1e-24}} \right)
		\left( \frac{d}{1 \mathrm{kpc}} \right)
			\left( \frac{\SI{100}{\hertz}}{f} \right)^3.
\end{equation}
Figure \ref{fig:rmodeULs} shows the upper limits on the $r$-mode amplitude $\alpha$ for different distances $d$.

\begin{figure}
\includegraphics[width=\columnwidth]{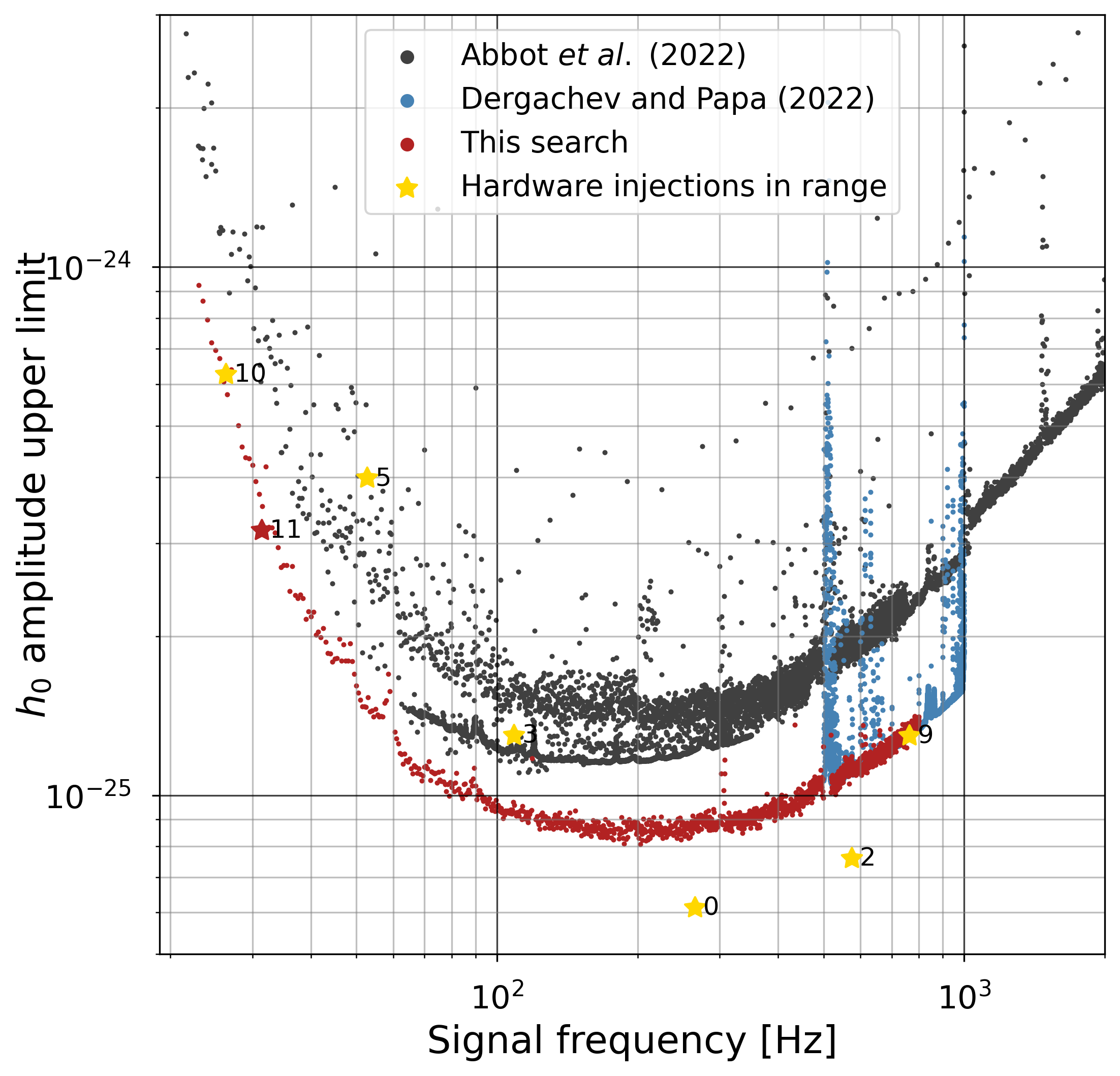}
\caption{The smallest gravitational-wave amplitude $h_0$ that we can exclude from our target population. We compare to other all-sky searches in LIGO O3 data \citep{KAGRA:2022dwb, Dergachev:2022lnt}. There are multiple curves for \citet{KAGRA:2022dwb} corresponding to multiple analysis pipelines. The golden stars are the gravitational-wave amplitudes $h_0$ of the hardware injections. Of the seven hardware injections in our search range, we recover all but hardware injection 11 (red star), due to its parameters being at a high mismatch point in the search grid. To miss one injection out of seven is perfectly consistent with our $90\%$ confidence upper limits.}
\label{fig:h0ULs}
\end{figure}

\begin{figure}
\includegraphics[width=\columnwidth]{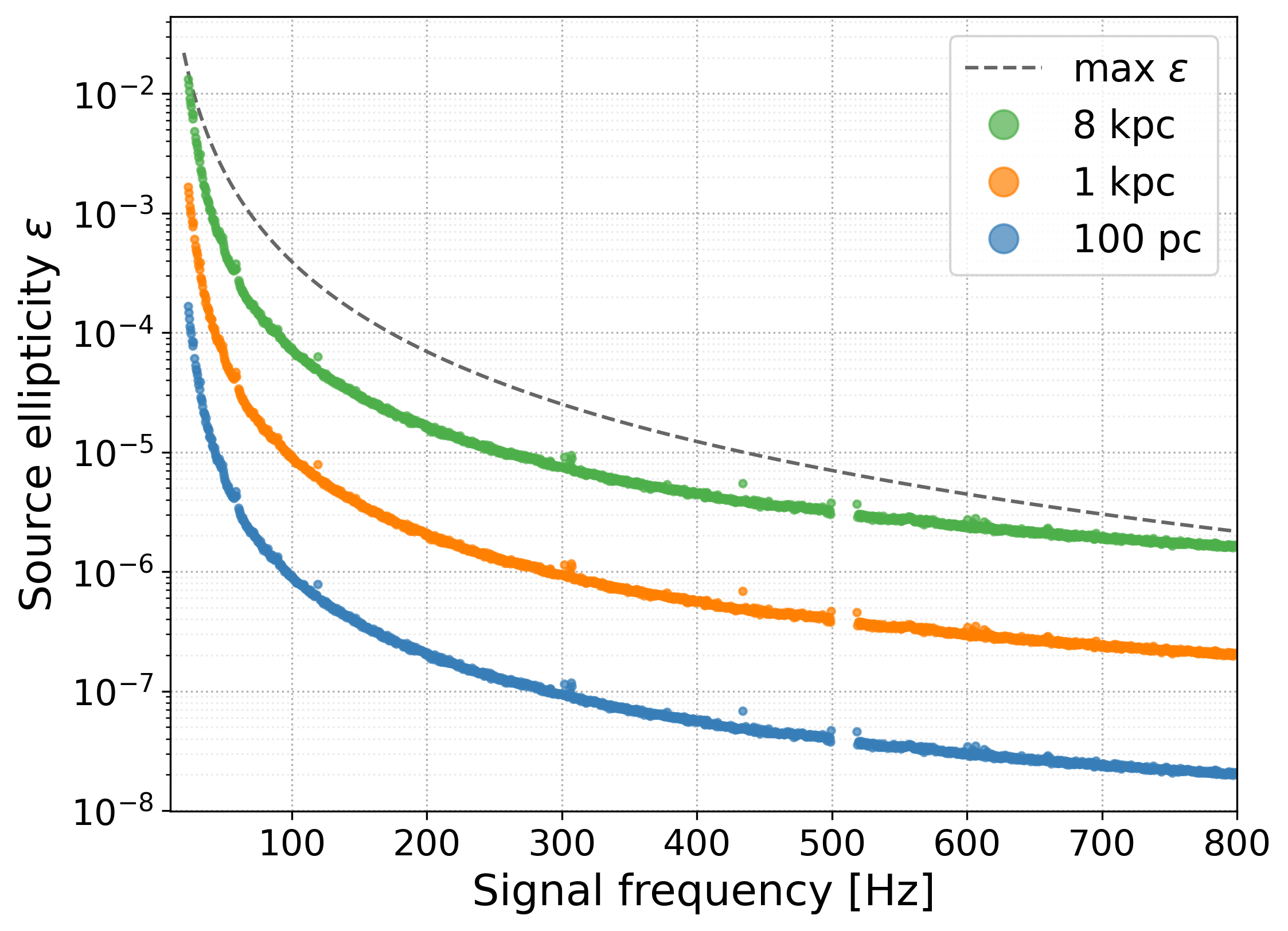}
\caption{Upper limits on the neutron star ellipticity $\varepsilon$ at different distances. The dashed line shows the maximum ellipticity probed due to the maximum spin-down of this search. The slightly increased upper limit values near 120 Hz, 305 Hz, and 435 Hz  -- also visible in the $h_0$ upper limit plot -- are due to decreased search sensitivity, due to the line-cleaning procedure.}
\label{fig:epsilonULs}
\end{figure}

\begin{figure}
	\includegraphics[width=\columnwidth]{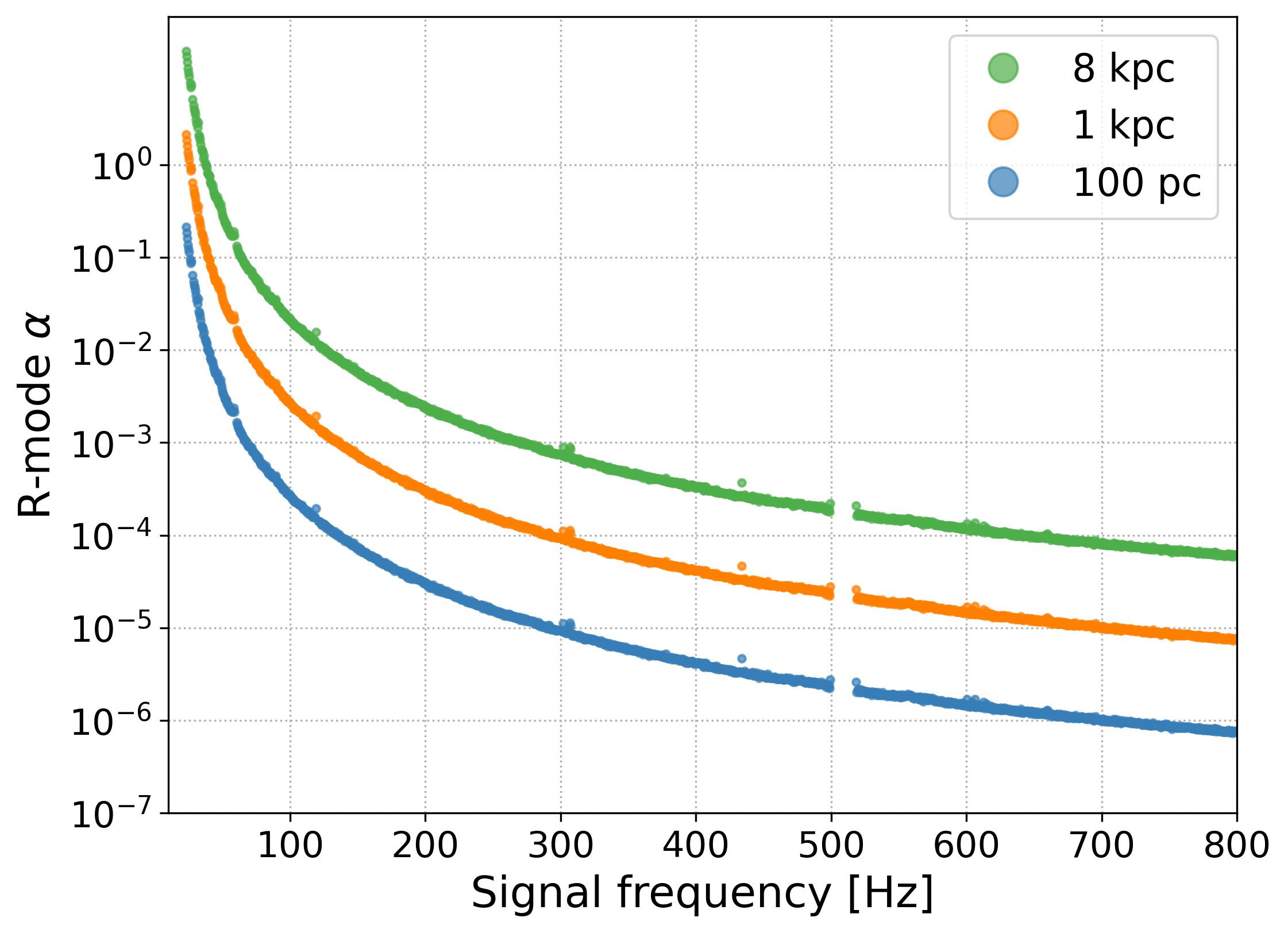}
	\caption{Upper limits on the $r$-mode amplitude $\alpha$ at different distances.}
	\label{fig:rmodeULs}
\end{figure}

\section{Conclusions}
\label{sec:conclusions}

We present the results from an \EaH{} all-sky search for continuous gravitational waves with frequency between $\SI{\paramfmin}{\hertz} \leq f \leq \SI{\paramfmax}{\hertz}$ and a spin-down of $\SI{\paramfdotmin}{\hertz\per\second} \leq \fdot \leq \SI{\paramfdotmax}{\hertz\per\second}$ in the LIGO O3 public data. 

Of the $\paramtotaltemplates$ waveforms searched, we identify the most promising 3.5 million and investigate them with a nine-stage hierarchical follow-up. The only surviving candidates are associated to the hardware injection fake signals.

We recover all hardware injections that lie above/at our upper limit level and even two with amplitude significantly below the upper limit curve. All signal parameters are recovered with high accuracy.

The most stringent 90\% confidence level upper limit on the gravitational-wave amplitude $h_0$ is placed at $\mostStringentULFreq$ Hz, at the level of $\mostStringentUL$. This excludes neutron stars with ellipticity $\varepsilon$, at a distance $d$ and rotating with periods $P$:
\begin{equation}
\label{eq:ULs}
P < 5~{\textrm{ms}} ~~~{\textrm{and}}~~~\varepsilon > 5\times 10^{-8} \left[ { {d}\over{ 100~{\textrm{pc}} } } \right].
\end{equation}

Depending on the assumptions, the nearest unknown neutron star could be as close as $\approx 10$ pc \citep{Dergachev:2020fli,Pagliaro2023}, where this search probes the interesting region of ellipticities between $10^{-7}$ and $10^{-9}$ for sufficiently fast-spinning ($\geq 50$ Hz) objects. \citet{Pagliaro2023}, however, find that the average distance of nearest neutron star {\it{spinning in band}} is approximately 100 pc. This means that, in practice, we are most likely probing ellipticities between $10^{-6}$ and $10^{-8}$.

It is hence natural to ask if objects with ellipticities larger than $10^{-8}$ could exist. There is much uncertainty about the maximum ellipticity that a neutron star crust can support, and the mechanisms to create such ellipticity, especially for isolated objects. While some models predict maximum ellipticities around $\sim10^{-5}$ \citep{JohnsonMcDaniel:2012wg, Morales:2022wxs}, other works show that the maximum may be only at $10^{-9}$ \citep{Bhattacharyya:2020paf,Gittins:2021zpv}. Overall, ellipticities at the level of $10^{-6}$ are usually considered reasonable, and the reach of this search for objects with this deformation and rotating faster than 150 Hz is 1 kpc.

Young, energetic neutron stars are promising emitters of gravitational waves due to $r$-modes. The amplitudes probed here are not implausible for a star a few years -- at most a few decades -- old, born anywhere in the Galaxy \citep{Bondarescu:2008qx}.

This is the most sensitive all-sky search performed on this parameter space to date. Still, we cannot claim that with the same computational budget, an even more sensitive search could not be performed. Rather, our search setup investigations did not identify one. Our investigations are resource-intensive, because they rely on Monte Carlo simulations aimed at {\it{measuring}} the detection efficiency of signals under different search conditions. And because they are resource-intensive, the scope of the considered setups is limited. We would like to overcome these limitations in the future. One approach is to make the simulations more efficient; alternatively, one could use an analytical model to predict the sensitivity of the search. Such a model would have to be accurate to a few percent, which is the scale of the sensitivity depth improvements amongst different methods working on the same data. Given the nonlinearity of many of the steps (clustering, to mention one), obtaining this accuracy in the prediction of the sensitivity is not trivial.

However the search setup will be arrived at, we look forward to the LIGO/Virgo data of the next observing run (O4), which we will use to probe an even more interesting range of deformations/distances.

\acknowledgments
We thank the \EaH\ volunteers, without the support of whom this search could not have happened. We thank Gianluca Pagliaro for useful discussions on the detectability of Galactic neutron stars. We acknowledge the use of Topcat \citep{2005ASPC..347...29T}. The Einstein@Home project is supported by the NSF award 1816904.
This research has made use of data or software obtained from the Gravitational Wave Open Science Center (gwosc.org), a service of LIGO Laboratory, the LIGO Scientific Collaboration, the Virgo Collaboration, and KAGRA. 

\newpage

\bibliography{paperBibApJ}{}
\bibliographystyle{aasjournal}
\bibstyle{aasjournal}

\end{document}